\documentclass[twocolumn]{aastex61}

\newcommand{\archain}{{\sc ar-chain}}
\newcommand{\ketju}{{\sc ketju}}
\newcommand{\gadget}{{\sc gadget-3}}

\graphicspath{{./}{figures/}}

\received{\today}
\revised{DD.MM.YYYY}
\accepted{DD.MM.YYYY}
\submitjournal{ApJL}

\shorttitle{Large cores and stellar kinematics in massive ETGs with SMBHs}
\shortauthors{Rantala et al.}

\begin{document}

\title{The simultaneous formation of cored, tangentially biased, and kinematically decoupled centers in massive early-type galaxies}


\correspondingauthor{Antti Rantala}
\email{antti.rantala@helsinki.fi}

\author[0000-0001-8789-2571]{Antti Rantala}
\affil{Department of Physics, University of Helsinki, Gustaf H\"allstr\"omin 
katu 2, 00560 Helsinki, Finland}

\author[0000-0001-8741-8263]{Peter H. Johansson}
\affiliation{Department of Physics, University of Helsinki, Gustaf 
H\"allstr\"omin katu 2, 00560 Helsinki, Finland}

\author[0000-0002-7314-2558]{Thorsten Naab}
\affiliation{Max-Planck-Institut f\"ur Astrophysik, Karl-Schwarzchild-Str. 1, 
D-85748, Garching, Germany}

\author[0000-0003-2868-9244]{Jens Thomas}
\affiliation{Max-Planck-Institut f\"ur Extraterrestriche Physik, 
Giessenbach-Str. 1, D-85741, Garching, Germany}

\author[0000-0002-0588-7259]{Matteo Frigo}
\affiliation{Max-Planck-Institut f\"ur Astrophysik, Karl-Schwarzchild-Str. 1, 
D-85748, Garching, Germany}

\begin{abstract}
We study the impact of merging supermassive black holes (SMBHs) on the central regions of massive early-type galaxies (ETGs) using a series of merger simulations with varying mass ratios. The ETG models include realistic stellar and dark matter components and are evolved with the \gadget{} based regularized tree code \ketju. We show that observed key properties of the nuclear stellar populations of massive ETGs, namely flat stellar density distributions (cores), tangentially biased velocity distributions and kinematically decoupled (counter-)rotation can naturally result from a single process $-$ the scouring by SMBHs. Major mergers with mass ratios of $q>1/3$ produce flat, tangentially biased cores with kinematically distinct components. These kinematic features originate from reversals of the SMBH orbits caused by gravitational torques after pericenter passages. Minor mergers ($q\lesssim1/3$) on the other hand, form non-rotating cores and the orbit reversal becomes less important. Low-density stellar cores scoured in (multiple) minor mergers are less tangentially biased. This implies that the nuclear stellar properties of massive ETGs can be solely explained by stellar dynamical processes during their final assembly without any need for `feedback' from accreting black holes. We predict a strong correlation between decoupled cores, central anisotropy and merger history: decoupled cores form in binary mergers and we predict them to occur in elliptical galaxies with the strongest central
anisotropy. Measurements of the central orbital structure are the key to understanding the number of mergers a given galaxy has experienced.

\end{abstract}


\section{Introduction}

The presence of supermassive black holes (SMBHs) is ubiquitous in all massive galaxies in the local Universe \citep{kormendy2013}. A significant fraction of massive early-type galaxies (ETGs) show flat density profiles \citep{Faber1997}, tangentially biased central velocity dispersions \citep{Thomas2014} and kinematic decoupling (e.g. \citealt{Emsellem2007,Ene2018}). In this Letter we present merger simulations of massive model ETGs with SMBHs and dark matter that demonstrate that all three features, in quantitative agreement with observations, can result simultaneously from the merging of supermassive black holes in massive galaxies. In addition, this scouring process will not affect the homogeneity of the stellar population properties in ETG centers, such as their stellar ages and metallicities.

Massive early-type galaxies ($M_\star \gtrsim 10^{10.8} M_\odot$) have photometric and stellar kinematic properties which differ from lower mass ETGs (e.g. \citealt{Kormendy2009ApJS}). Their stellar components typically show slow or no global rotation \citep{Emsellem2011}. In addition, most massive ETGs have nearly flat central light profiles (cores) over a region that scales with the mass of the central black hole (\citealt{Lauer2007a, Rusli2013b}) and its sphere-of-influence \citep{Thomas2016}. About $\sim 80 \%$ of nearby massive and slowly rotating galaxies have cores \citep{Krajnovic2013} and the fraction of slow rotators rises rapidly with increasing stellar mass \citep{Veale2017b}. Furthermore, detailed orbit models from the SINFONI black hole survey revealed a very uniform orbital structure near the black hole, with a predominance of tangential over radial motions \citep{Thomas2014}.

Theoretically the above features can be understood in a galaxy merger framework (\citealt{Naab2017} ). Massive ETGs are gas-poor systems and their late ($z<2$) cosmological assembly is dominated by a phase of gas-poor (dry) merging, in which the ETGs accrete stars formed mainly in progenitors outside the main galaxy (e.g. \citealt{DeLucia2007, Oser2010, 2012Johansson, 2015Wellons, rodriguez-gomez2016}). This provides plausible explanations for the observed rapid size growth \citep{Naab2009}, the high S\'ersic indices of the surface brightness profiles \citep{Hilz2013}, and the slow rotation \citep{Hernquist1991, Naab2014}. 

The formation of low density cores and the observed velocity anisotropy can be explained by the scouring of stars in the centers by merging supermassive black holes (e.g. \citealt{Milosavljevic2001, Milosavljevic2003}). Recent equal-mass merger simulations with more realistic multi-component ETG models including supermassive black holes and dark matter halos have shown that the cores become larger and the anisotropy is stronger for more massive black holes, in good agreement with observations \citep{Rantala2018}.

 About half of the massive ETGs with cores show kinematic decoupling or counter rotation in their central region \citep{Krajnovic2013}. In addition, the observed high stellar metallicities and old stellar ages are similar to the surrounding stellar populations (e.g. \citealt{Bender1994}). Galaxy merger simulations have early on been able to explain the formation of counter rotating cores (e.g. \citealt{Barnes1992}). However, the process nearly always require gas dissipation and counter-rotating retrograde merger orbits in order to produce central kinematic decoupling (e.g. \citealt{Balcells1990,Jesseit2007,Bois2011}).
 
A recent study by \cite{Tsatsi2015} reported a kpc-scale counter-rotating old stellar component in a prograde simulation of merging gas-rich disks. They attributed this formation process to reactive forces during the merger that cause a short-lived change in the orbital spins of the progenitor galaxies. However, gas dissipation in galaxy centers and the resulting star formation results in even denser power-law density profiles and no cores (e.g. \citealt{Barnes1991,Hopkins2009}). In addition, it seems difficult to construct a fine tuned scenario where AGN feedback, acting at the right time, would allow for the formation of kinematically decoupled cores (KDC) and still expel the gas to prevent the formation of a density cusps (e.g. \citealt{Capelo2017, Frigo2018}). A plausible scenario for the formation of tangentially biased and decoupled cores is therefore still missing. 

In this letter we present a dissipationless merger scenario, which allows for the simultaneous formation of tangentially biased and kinematically decoupled cores with homogeneous stellar population properties.

\begin{table*}
\begin{center}
\begin{tabular}{ | c | c | c | c | c | c | c |}
    \hline
    \multicolumn{7}{|c|}{Initial conditions}\\
    \hline
    Progenitor & $M_\mathrm{DM}$ [$10^{13} M_\odot$] & $M_\star$ [$10^{10} M_\odot$] & $M_\bullet$ [$10^{9} M_\odot$] & $R_\mathrm{e}$ [kpc] & $N_\mathrm{DM}$ [$\times 10^6$] & $N_\star$ [$\times 10^6$]\\ 
    \hline
    IC-1 & 1.50 & 8.30 & 1.70 & 3.50 & 2.00 & 0.83\\
    \hline
    IC-2 & 1.88 & 10.34 & 2.13 & 4.16 & 2.50 & 1.04\\
    \hline
    IC-3 & 2.50 & 13.83 & 2.93 & 4.95 & 3.33 & 1.38\\
    \hline
    IC-4 & 3.75 & 20.75 & 4.25 & 5.90 & 5.00 & 2.08\\
    \hline
    IC-5 & 7.50 & 41.50 & 8.50 & 7.00 & 10.00 & 4.15\\
    \hline
    IC-5-nobh & 7.50 & 41.50 & - & 7.00 & 10.00 & 4.15\\
    \hline
    \hline
    \multicolumn{7}{|c|}{Binary mergers}\\
    \hline
    Merger remnant &  $M_\mathrm{DM}$ [$10^{13} M_\odot$] & $M_\star$ [$10^{10} M_\odot$] & $M_\bullet$ [$10^{9} M_\odot$] & \multicolumn{2}{|c|}{Merged progenitors} & Mass ratio\\
    \hline
    ETG-5-1 & 9.00 & 49.80 & 10.20 &  \multicolumn{2}{|c|}{  IC-5 + IC-1} & 5:1 \\
    \hline
    ETG-4-1 & 9.38 & 51.88 & 10.63 & \multicolumn{2}{|c|}{ IC-5 + IC-2} & 4:1\\
    \hline
    ETG-3-1 & 10.00 & 55.33 & 11.33 &  \multicolumn{2}{|c|}{ IC-5 + IC-3} & 3:1\\
    \hline
    ETG-2-1 & 11.25 & 62.25 & 12.75 & \multicolumn{2}{|c|}{ IC-5 + IC-4} & 2:1\\
    \hline
    ETG-1-1 & 15.00 & 83.00 & 17.00 & \multicolumn{2}{|c|}{ IC-5 + IC-5} & 1:1\\
    \hline
    ETG-1-1-nobh & 15.00 & 83.00 & - & \multicolumn{2}{|c|}{ IC-5-nobh + IC-5-nobh} & 1:1\\
    \hline
    \hline
    \multicolumn{7}{|c|}{Multiple merger generations}\\
    \hline
    Merger remnant &  $M_\mathrm{DM}$ [$10^{13} M_\odot$] & $M_\star$ [$10^{10} M_\odot$] & $M_\bullet$ [$10^{9} M_\odot$] & \multicolumn{2}{|c|}{Merged progenitors} & Mass ratio\\
    \hline
    ETG-M-A & 30.00 & 166.00 & 34.00 & \multicolumn{2}{|c|}{ ETG-1-1 + ETG-1-1} & 1:1\\
    \hline
    ETG-M-B & 15.00 & 83.00 & 17.00 & \multicolumn{2}{|c|}{ IC-5 + $5\times$ IC-1} & 5:1 - 9:1\\
    \hline
    \end{tabular}
    
\caption{The properties of the progenitor galaxies and merger remnants in this study.}
\label{table: t1}
\end{center}
\end{table*}

\section{Numerical simulations}

\subsection{The \ketju{} code}

The simulations are run using the \ketju{} code \citep{Rantala2017}, an extension of the widely used tree-SPH simulation code \gadget{} \citep{Springel2005}. The key feature of the code is the inclusion of a regularized region around every SMBH in a simulation. The non-softened gravitational dynamics of systems of SMBHs and stars around any SMBH ($r<r_\mathrm{chain}$) in the simulation is computed using the regularized \archain{} \citep{Mikkola2008} integrator while the dynamics of the remaining particles is computed with the \gadget{} leapfrog using 
the tree force calculation method (\citealt{Rantala2017, Rantala2018}). The combination of the regularized integrator and \gadget{} enables simulations with high particle numbers and accurate small-scale SMBH dynamics.

As in \cite{Rantala2018} we set $r_\mathrm{chain} = 10$ pc for the size of the subsystems. The gravitational softening lengths are set to $\epsilon_\mathrm{DM} = 100$ pc and $\epsilon_\star = 3.5$ pc for the dark matter (DM) and stellar particles, respectively. Particles within $r<r_\mathrm{pert}$ of any SMBH act as near-field perturbers of the regularized subsystem, with $r_\mathrm{pert} = 2\times r_\mathrm{chain}$, whereas the more distant particles act as far-field perturbers. The \gadget{} integrator accuracy and the standard tree cell opening criterion parameter are set to $\eta = 0.002$ and $\alpha = 0.005$. We set the \archain{} GBS tolerance to $\eta_\mathrm{GBS} = 10^{-6}$. Post-Newtonian corrections up to order PN3.5 are used in the equations of motion of the SMBHs.

\subsection{Initial conditions}

The progenitor galaxies consist of a Dehnen \citep{Dehnen1993} stellar component ($\gamma = 3/2$), a Hernquist ($\gamma = 1$) dark matter halo and a central SMBH. The isotropic velocity profiles of the progenitors are constructed using the distribution function method and Eddington's formula \citep{binney2008}. The parameters ($M_\mathrm{DM},M_\star,M_\bullet,R_\mathrm{e}$) of the progenitor galaxies are presented in Table \ref{table: t1}. 

The most massive progenitor galaxy IC-5 is identical to the IC `$\gamma$-1.5-BH-6' of \cite{Rantala2018}. The progenitors IC-1 - IC-4 are scaled-down versions of IC-5, the masses of the components divided by a factor of $2$ to $5$. The effective radius of the least massive progenitor IC-1 is smaller by a factor of two compared to the most massive progenitor (IC-5) \citep{Hilz2012} and the progenitors IC-2, IC-3 and IC-4 have effective radii intermediate between these two values following a power-law scaling. The dark matter scale radius $a_\mathrm{DM}$ is chosen to set the dark matter fraction to $f_\mathrm{DM} = 0.25$ within $R_\mathrm{e}$ \citep{Rantala2018}. The particle masses are $m_\mathrm{DM} = 7.5\times10^6 M_\odot$ and $m_\star = 1.0\times10^5 M_\odot$.

\begin{figure*}
\includegraphics[width=\textwidth]{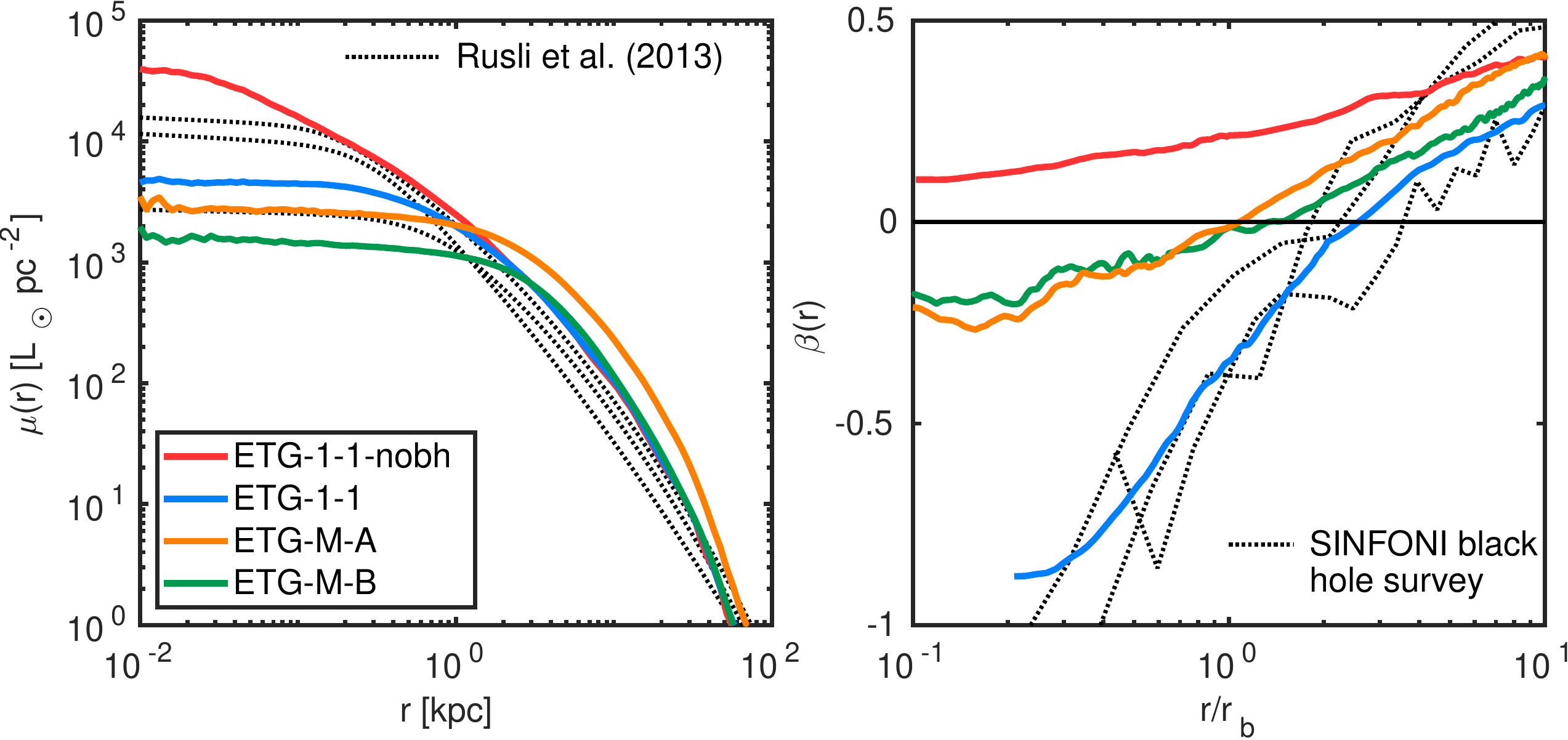}
\caption{Left panel: the surface brightness profiles $\mu(r)$ of four representative merger remnants, $M/L_\star$ = 4.0 \citep{Thomas2016}. Observed surface brightness profiles from \cite{Rusli2013b} are shown as dotted lines. The core-S\'ersic fits yield core radii of $r_\mathrm{b}=0.49$ kpc (ETG-1-1), $r_\mathrm{b}=1.51$ kpc (ETG-M-A) and $r_\mathrm{b}=1.55$ kpc (ETG-M-B), respectively. Right panel: the stellar velocity anisotropy profiles of the remnants with observed profiles of core ellipticals (dashed black lines) \citep{Saglia2016}. For the run without SMBHs (red line) we set $r_\mathrm{b} = 1$ kpc. Binary major mergers without SMBHs yield radially biased anisotropy profiles while the inclusion of the SMBHs (blue line) produces a strongly tangentially biased velocity distribution in the core region. Repeated mergers (green and orange lines) render the central velocity distribution closer to isotropic.}
\label{fig1}
\end{figure*}

\begin{figure*}
\includegraphics[width=\textwidth]{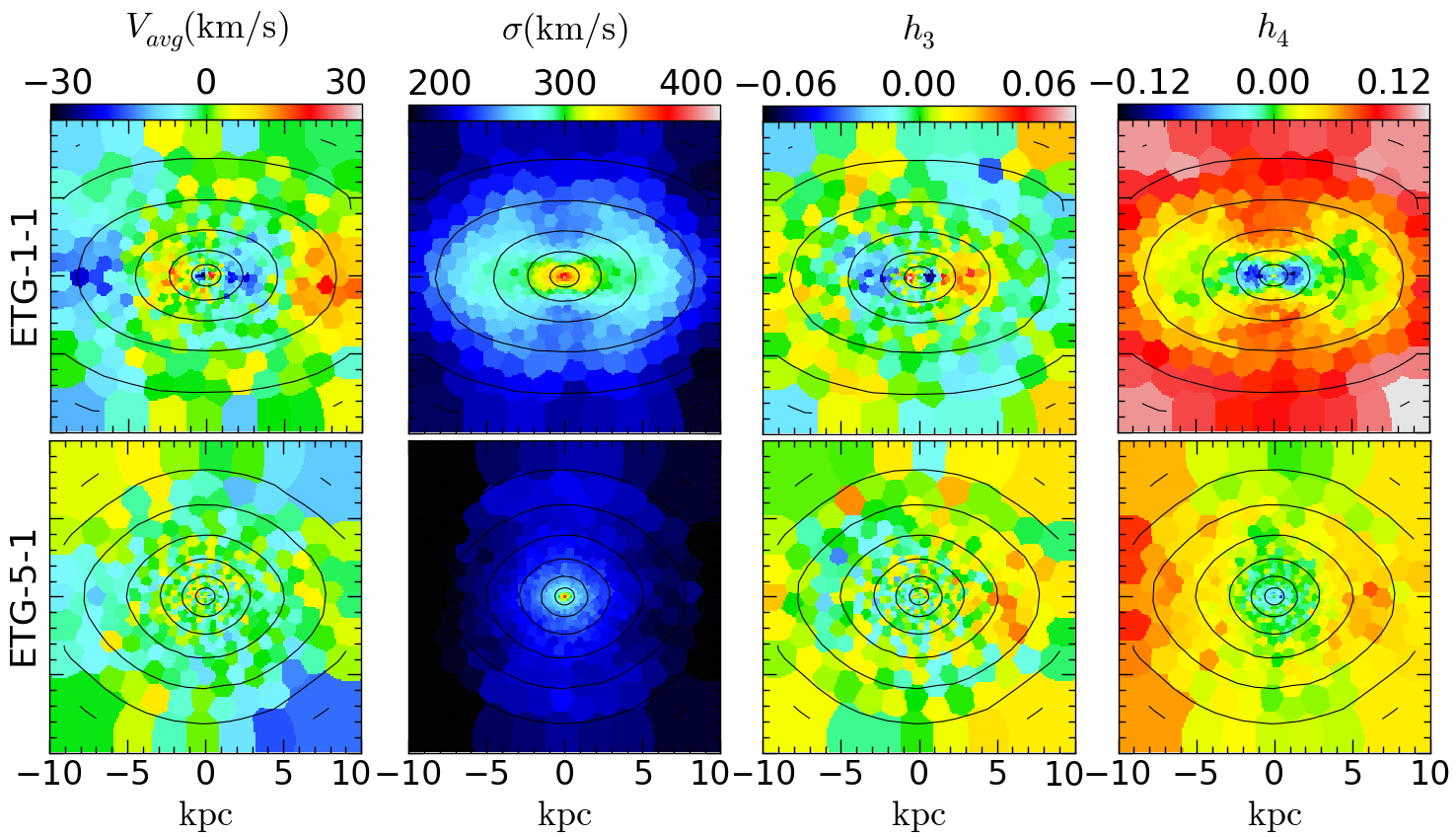}
\caption{The 2D kinematic maps of the merger remnants ETG-1-1 (top row) and ETG-5-1 (bottom row) presenting the mean LOS velocity, velocity dispersion and Gauss-Hermite moments $h_3$ and $h_4$. The binary major merger remnant ETG-1-1 shows complex central rotation signatures with an anticorrelation of $V_\mathrm{avg}$ and $h_3$. The minor merger remnant ETG-5-1} does not show such rotation features.
\label{fig2}
\end{figure*}

For the base simulation we chose the $\gamma$-1.5-BH-6 run from \cite{Rantala2018} with a NGC 1600-like SMBH mass, which is somewhat above the host galaxy-BH scaling relations. We simulate five binary galaxy mergers including SMBH with mass ratios of 1:1, 2:1, 3:1, 4:1 and 5:1 (merger remnants ETG-5-1 to ETG-1-1). The 1:1 merger is additionally run without SMBHs (ETG-1-1-nobh). We continue the 5:1 merger run with subsequent merger generations until five minor mergers are completed (ETG-M-B) \citep{Hilz2013}. We remerge remnant ETG-1-1 with an identical remnant to study whether the physical features (core, anisotropy, rotation) formed in the first merger disappear in a second major merger (ETG-M-A). The eight merger remnants are listed in the middle and bottom panels of Table \ref{table: t1}. 

All merger orbits are nearly parabolic the pericenter distance being $r_\mathrm{p} \sim 0.5\times R_\mathrm{e}$ of the host galaxy. After each minor merger, the merger remnant is reoriented so that the satellite galaxies fall in from random directions with respect to the principal axis of the host. All simulations were run until the GW-driven final SMBH merger, with the maximum merger timescale being $\sim 4$ Gyr.

\section{Results}

\subsection{Surface brightness profiles, mass deficits and velocity anisotropies}

The surface brightness profiles of the merger remnants ETG-1-1, ETG-1-1-nobh, ETG-M-A and ETG-M-B are presented in Fig. \ref{fig1}. The observed surface brightness profiles of three ellipticals with large cores (NGC 1600 and NGC 4472, NGC 5328) from \cite{Rusli2013b} are shown for a qualitative comparison. Most simulated merger remnants have lower central surface brightness compared to the observed galaxies. We attribute this quantitative difference to the fact that our simulations contain very massive SMBH binaries, with masses well above the typical values inferred from the SMBH - host galaxy relation (\citealt{Thomas2016, Rantala2018}).

The final stellar and SMBH masses of the remnants ETG-1-1, ETG-1-1-nobh and ETG-M-B are equal. We compute the central mass deficits from the surface brightness profiles in the core region ($r<r_\mathrm{b}$) of the two merger remnants with SMBHs with respect to the non-scoured remnant ETG-1-1-nobh. Repeated minor mergers (ETG-M-B) result in a higher mass deficits. In five consecutive minor mergers, the $M_\mathrm{def}$ increases as $M_\mathrm{def}/M_\bullet =$ $0.58$ ($0.50$), $1.55$ ($1.00$), $2.46$ ($1.50$), $3.08$ ($2.00$), $3.45$ ($2.50$). The values in the parenthesis are estimates by \cite{Merritt2006}: $M_\mathrm{def}/M_\bullet \sim 0.5 \mathcal{N}$ in which $\mathcal{N}$ is the number of mergers. Our somewhat higher mass deficits are explained by the fact that \cite{Merritt2006} measures the deficits when the binaries become hard whereas we follow the binary evolution to smaller separations.

The velocity anisotropy profiles, $\beta(r) = 1-\sigma_\mathrm{t}^2/2\sigma_\mathrm{r}^2$, of the four simulated mergers remnants are shown in Fig. \ref{fig1} with three observed galaxies (\citealt{Thomas2014, Saglia2016}). The profiles are computed in radial bins with a bin width of $0.2$ kpc. The anisotropy profile of ETG-1-1-nobh is radially biased ($\beta > 0$) in the absence of SMBH binary. As in \cite{Rantala2018}, the remnant ETG-1-1 is strongly tangentially biased ($\beta < 0$) in the core region, being in good agreement with the velocity anisotropy profiles of the observed core galaxies. The repeated mergers produce remnants ETG-M-A and ETG-M-B which are mildly tangentially biased in the central region, with each merger generation rendering the velocity anisotropy profile of the core region closer to isotropic.

The core sizes of the merger remnants built up in multiple mergers are $r_\mathrm{b} \sim 1.5 $ kpc, measured using the core-S\'ersic fit (\citealt{Graham2003}). This is in contrast to the cores formed in a single binary merger for which we find $r_\mathrm{b} \lesssim 0.5$ kpc. As the SMBH of the simulated remnant ETG-1-1 ($r_\mathrm{b} \sim 0.5$ kpc) is at the upper end of the observed SMBH masses, $M_\bullet = 1.7\times 10^{10} M_\odot$, it seems probable that very large cores ($r_\mathrm{b} \sim $ $3$-$4$ kpc) are formed more gradually by a large number of minor mergers with SMBHs (\citealt{Merritt2006}).

Simulations focusing on SMBH binaries in isolated galaxies or equal-mass binary mergers have connected flat low-density stellar cores in ETGs to central tangentially biased velocity anisotropies (\citealt{Quinlan1997,  Milosavljevic2001, Rantala2018}). Our results demonstrate that the formation of a stellar core by SMBH binary scouring produces a significantly less tangentially biased orbit distribution if the core was formed in a series of minor mergers instead of a single major merger. For individual galaxies, orbit modelling is the key to understanding their orbital structure and merger history \citep{Thomas2014}. Decoupled kinematic features are projection dependent and will only be observable at appropriate viewing angles. 

\subsection{Stellar kinematics}\label{section: 3.2}

\begin{figure}
\includegraphics[width=\linewidth]{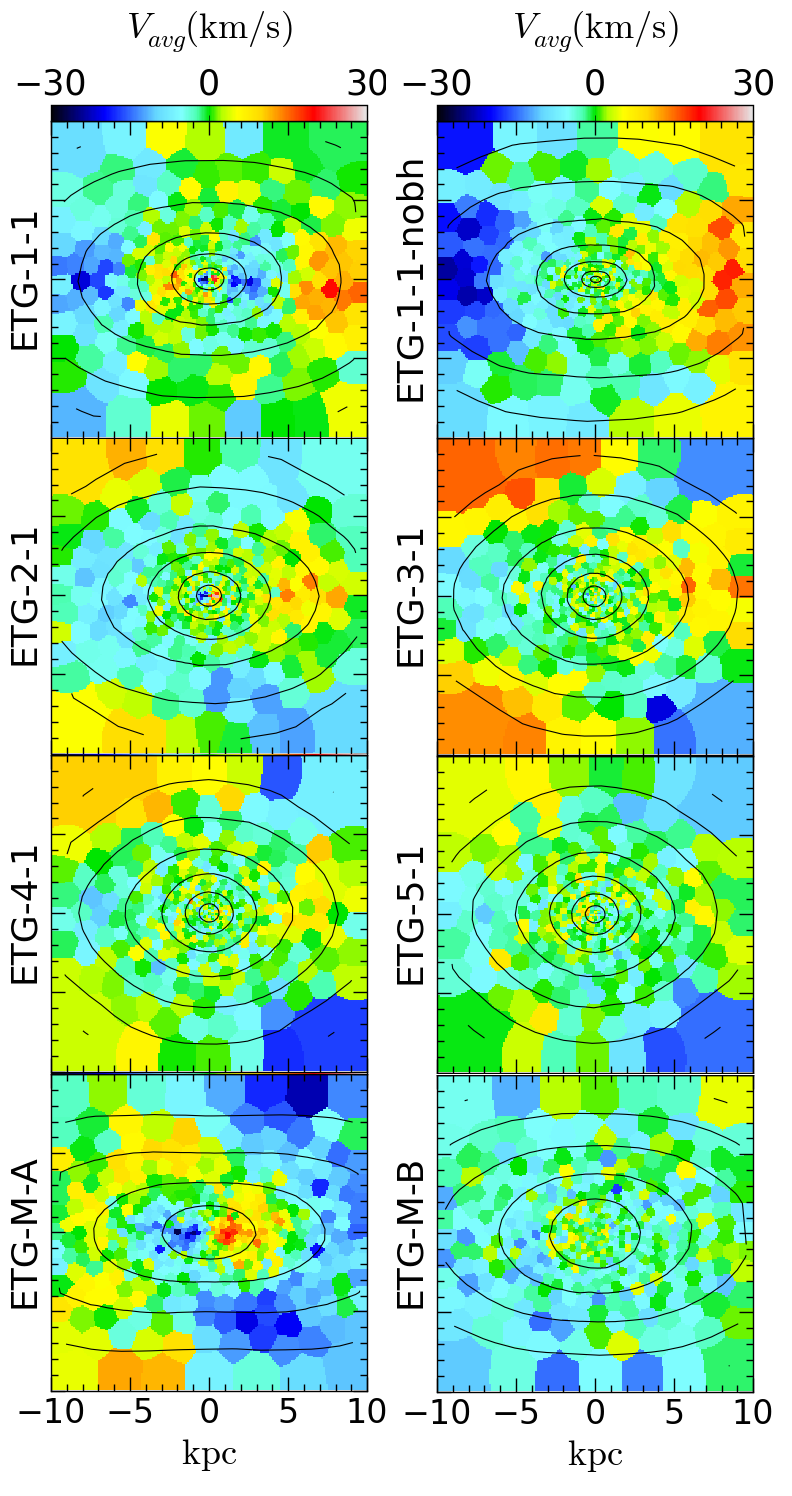}
\caption{The 2D kinematic maps of average LOS velocities of the merger remnants. Binary major mergers with SMBHs produce complex, decoupled kinematic regions at the central region. These features are weak or absent in minor mergers with SMBHs and in simulations without SMBHs.}
\label{fig3}
\end{figure}

Fig. \ref{fig2} presents the 2D kinematic maps of the merger remnants ETG-1-1 and ETG-5-1. The 2D maps are divided into Voronoi spaxels with an equal signal-to-noise ratio in each spaxel (\citealt{Cappellari2003, Naab2014}). The $V_\mathrm{avg}$, $\sigma$, $h_3$ and $h_4$ are computed in each spaxel by fitting a Gauss-Hermite function to the histogram of LOS velocities \citep{vanderMarel1993}. The galaxies are oriented using the reduced inertia tensor, the major axis lying on the x-axis of the panels. The surface brightness data is overlaid as black contour lines with a spacing of 1 magnitude. We present the mean LOS velocity maps of all the 8 simulated merger remnants in Fig. \ref{fig3}.

We characterize the rotation of the merger remnants using the spin parameter - ellipticity ($\epsilon$,$\lambda_\mathrm{R}$) criterion of \cite{Emsellem2011}. We find that the spin parameter is small for every remnant, $\lambda_\mathrm{R} < 0.03$ while the ellipticities are moderate, typically in the range $0.4<\epsilon<0.5$. All of our simulated merger remnants are slow rotators.

Decoupled central rotation is seen in remnants ETG-1-1, ETG-2-1 and ETG-M-A, all harboring SMBH binaries with mass ratios of $q = M_2/M_1 \gtrsim 0.5$. Especially the major merger remnant ETG-1-1 shows multiple decoupled, rotating kinematic subsystems, indicated by the anticorrelation of $V_\mathrm{avg}$ and $h_3$ (e.g. \citealt{Bender1994}). The maximum rotation velocity is $\sim 30$-$40$ km/s. The mean velocity maps of minor merger remnants formed in ETGs 3-1, 4-1, 5-1 and M-B either show slow rotation at $>5$ kpc or are completely featureless. The remnant ETG-1-1-nobh shows clear rotation $5$-$10$ kpc from the center while the central region itself is non-rotating. 

We use the kinemetry method of \cite{Krajnovic2011} to analyze the rotation features of the merger remnants. All our merger remnants are non-regular rotators (NRR) as deviations from simple rotation are significant. Using the kinematic types of \cite{Krajnovic2011} we classify ETG-4-1, ETG-5-1 and ETG-M-B as type NRR-LV (low rotation velocity, $< 5$ km/s). ETG-1, ETG-2 and ETG-M-A harbor counter-rotating cores (CRC). The classification of ETG-3-1 and ETG-1-1-nobh is somewhat spurious as the outer parts of the galaxies are rotating while the core regions are not.

\subsection{Origin of the rotation features in core ellipticals}

\begin{figure*}
\includegraphics[width=\textwidth]{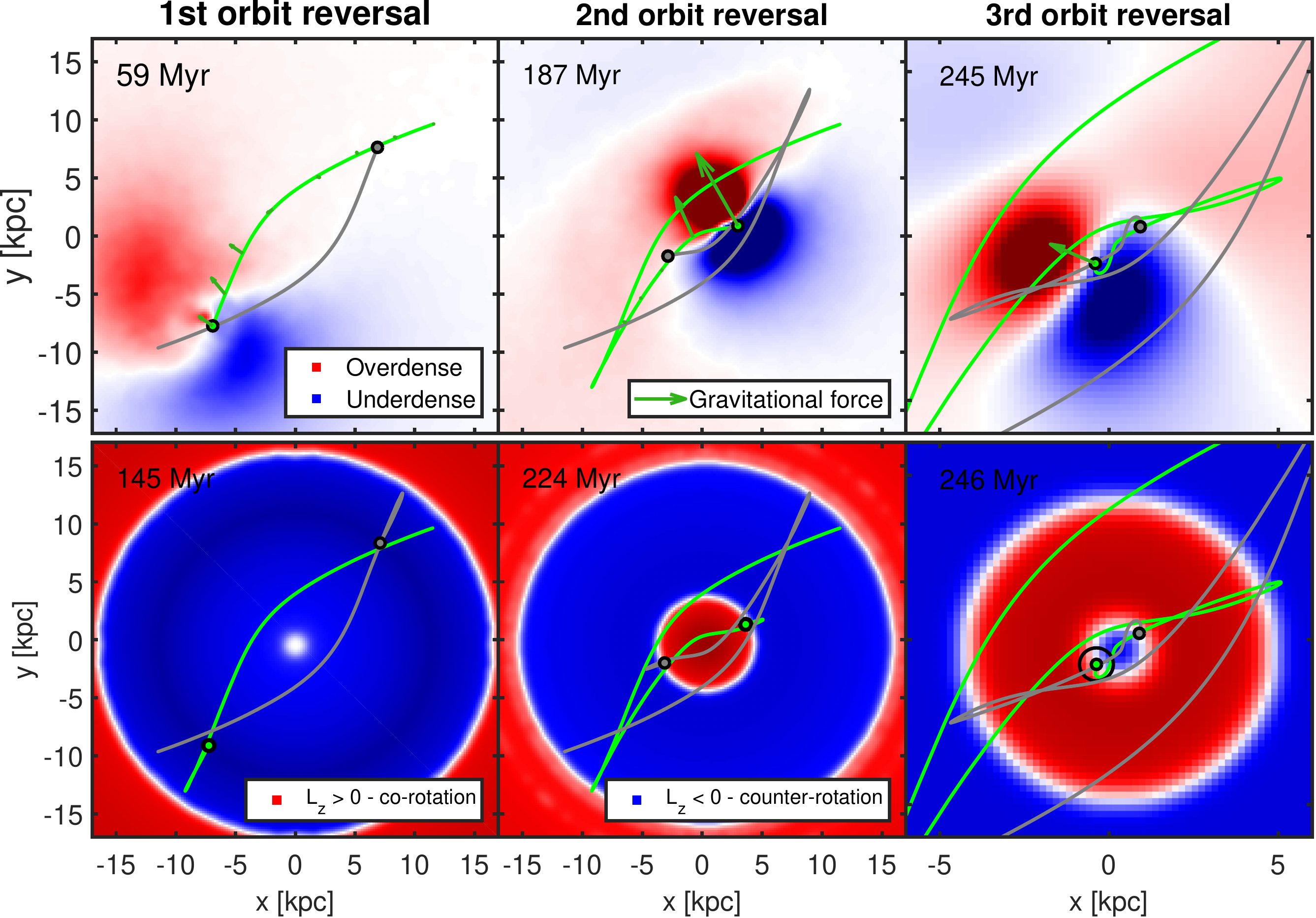}
\caption{Top panels: the reversals of the merger orbit in the run ETG-1-1. The panels show the SMBH orbits (green and gray), the stellar overdense and underdense regions (single progenitor), and the gravitational force vector from the tidal debris on one of the SMBHs. The vector is scaled down by a factor of 10 smaller in the right panel. Bottom panels: the azimuthally averaged stellar rotation direction (both progenitors). The apocenter distances determine where the average stellar rotation direction changes sign. The influence radius $r_\mathrm{infl}$ of a single SMBH is marked with a black circle in the bottom panels, but is visible only in the right zoom-in panel as in the other panels it is smaller than the SMBH symbol.}
\label{fig4}
\end{figure*}

In Fig. \ref{fig4} we show the origin of the complex rotation features of the major merger remnant ETG-1-1. In this simulation, the relative angular momentum of the SMBHs changes its sign three times during the galaxy merger. We relate the orbit reversals to two physical processes. \cite{Barnes2016} described a major merger process in which two bulges have a close encounter and are deflected from the major axis connecting their dark halos. The resulting gravitational torque reverses the relative angular momentum of the bulges. Another important effect is 'tidal self-friction' described in \cite{vandenBosch2018}, in which the center of the subhalo is strongly pulled to the direction of the tidally expelled material, resulting in the reversal of the orbit. 

The angular momentum reversals occur after the pericenter passages of the galaxies, before the apocenter. The encounter expels tidal debris from the galaxies, which pulls the centers of the galaxies to the direction of the stellar overdensity. Both dark matter and stars are important in the first orbit reversal, but later the effect of dark matter becomes negligible. The apocenter distances of the progenitors after each flip correspond to the locations where the stellar angular momentum and the mean LOS velocity in the kinematic maps change their signs. This connects the large-scale merger orbits to the complex velocity features observed in the LOSVD maps.

We performed tests in which the other galaxy was replaced by a static potential, and the orbit reversals occurred in these simulations as well. We also varied the pericenter distance by factors of $0.5$-$2$ and the initial velocity by a factor of two. Each of these merger orbits had at least one orbit reversal. However, we stress that a more detailed study of the progenitor and merger orbit parameters is still required in order to assess how generic orbital reversals are in these types of merger simulations.

In Fig. \ref{fig3} the outer rotation features are only seen in binary major mergers. \cite{Rantala2018} showed that a more centrally concentrated initial stellar density profile produces more strongly rotating outer regions, and that the central decoupled region becomes more prominent with increasing initial SMBH mass. Comparing the velocity maps of ETG-1-1, ETG-1-1-nobh, ETG-2-1 and ETG-3-1, the formation of the innermost decoupled region requires the presence of a close to equal-mass SMBH binary.

For less concentrated progenitor galaxies, smaller SMBH masses and large merger mass ratios the rotation features are weak or short-lived. Our simulations demonstrate that the binary mass ratios larger that approximately $q \gtrsim 1/3$ are needed to form long-lived central, kinematically decoupled regions in dry mergers of massive early-type galaxies.

\section{Conclusions}

The binary major galaxy mergers presented in this study allow for the simultaneous formation of tangentially biased and kinematically distinct low-density cores in massive early-type galaxies. Earlier formation models of these objects have relied on dissipational effects or on the merger of rotating disk galaxies on prograde orbits in order to produce kinematically distinct central regions. However, forming an anisotropic low-density core with kinematically decoupled regions has been proven to be difficult in these formation scenarios. Our progenitor galaxies are gas-free and initially non-rotating. Central kinematically distinct regions originate from the orbit reversals during the galaxy mergers.

The orbit reversals leave an imprint on the LOS velocity distributions of the merger remnants in the case of binary major mergers with mass ratios of $q \gtrsim 1/3$. Cuspy initial stellar profiles and more massive SMBHs lead to stronger central rotation signatures. The locations of the rotation reversals in the 2D kinematic maps correspond to the apocenters of the merger orbits after each orbit reversal (see \citealp{Tsatsi2015}). We also predict that, on average, galaxies with kinematically decoupled cores have a stronger central tangential anisotropy than galaxies without core rotation.

We find complex kinematic features in our cored merger remnants with mean LOS velocities in the range of $10$-$40$ km/s on spatial scales of up to $\sim 20$ core radii. These features are projection dependent, but our results demonstrate the feasibility of using rotation features as a first-order statistical tool to constrain the formation histories of massive, slowly-rotating, cored ETGs. 

Finally, we have demonstrated that the formation of a low-density, flat central core in an ETG by SMBH binary scouring produces less tangential orbit anisotropy near the center if the core was scoured in a series of minor mergers with SMBHs. Our results stress that precise measurements of black-hole masses and of the central orbital structure, e.g. through Schwarzschild orbit superposition models, is the key to understanding the merger history of ETGs and their SMBHs.

\small
\begin{acknowledgements}
The authors thank Simon White and Frank van den Bosch for discussions and valuable input. The numerical simulations were performed on facilities hosted by the CSC -IT Center for Science in Espoo, Finland. A. R. acknowledges support from the MPA Garching Visitor Programme.  P. H. J. and A. R. acknowledge the support of the Academy of Finland grant 274931. T. N. acknowledges support from the DFG Cluster of Excellence 'Origin and Structure of the Universe'.
\end{acknowledgements}

\end{document}